\documentclass[aps,prl,twocolumn,showpacs,superscriptaddress]{revtex4-1}  
\usepackage{graphicx}  
\usepackage{dcolumn}   
\usepackage{amsmath}        
\usepackage{amssymb}   

\def\mj{M{\sc ajo\-ra\-na}}
\def\dem{D{\sc e\-mon\-strat\-or}}
\def\nbb{0$\nu\beta\beta$}

\begin{document}


\title{New Limits on Bosonic Dark Matter, Solar Axions, Pauli Exclusion Principle Violation, and Electron Decay from the \mj\ \dem} 
%
\newcommand{\blhill}{Department of Physics, Black Hills State University, Spearfish, South Dakota 57799, USA}
\newcommand{\ITEP}{National Research Center ``Kurchatov Institute'' Institute for Theoretical and Experimental Physics, Moscow 117218, Russia}
\newcommand{\JINR}{Joint Institute for Nuclear Research, Dubna 141980, Russia}
\newcommand{\lbnl}{Nuclear Science Division, Lawrence Berkeley National Laboratory, Berkeley, California 94720, USA}
\newcommand{\lanl}{Los Alamos National Laboratory, Los Alamos, New Mexico 87545, USA}
\newcommand{\queens}{Department of Physics, Engineering Physics and Astronomy, Queen's University, Kingston, Ontario K7L 3N6, Canada} 
\newcommand{\uw}{Center for Experimental Nuclear Physics and Astrophysics, 
and Department of Physics, University of Washington, Seattle, Washington 98195, USA}
\newcommand{\unc}{Department of Physics and Astronomy, University of North Carolina, Chapel Hill, North Carolina 27599, USA}
\newcommand{\duke}{Department of Physics, Duke University, Durham, North Carolina 27708, USA}
\newcommand{\ncsu}{Department of Physics, North Carolina State University, Raleigh, North Carolina 27695, USA}	
\newcommand{\ornl}{Oak Ridge National Laboratory, Oak Ridge, Tennessee 37831, USA}
\newcommand{\ou}{Research Center for Nuclear Physics, Osaka University, Ibaraki, Osaka 567-0047, Japan}
\newcommand{\pnnl}{Pacific Northwest National Laboratory, Richland, Washington 99352, USA}
\newcommand{\princeton}{Department of Physics, Princeton University, Princeton, New Jersey 08544, USA}
\newcommand{\ttu}{Tennessee Tech University, Cookeville, Tennessee 38505, USA}
\newcommand{\sdsmt}{South Dakota School of Mines and Technology, Rapid City, South Dakota 57701, USA}
\newcommand{\usc}{Department of Physics and Astronomy, University of South Carolina, Columbia, South Carolina 29208, USA}
\newcommand{\usd}{Department of Physics, University of South Dakota, Vermillion, South Dakota 57069, USA} 
\newcommand{\ut}{Department of Physics and Astronomy, University of Tennessee, Knoxville, Tennessee 37996, USA}
\newcommand{\tunl}{Triangle Universities Nuclear Laboratory, Durham, North Carolina 27708, USA}
\newcommand{\mpi}{Max-Planck-Institut f\"{u}r Physik, M\"{u}nchen 80805, Germany}
\newcommand{\tum}{Physik Department and Excellence Cluster Universe, Technische Universit\"{a}t, M\"{u}nchen 80805, Germany}


\affiliation{\lbnl}
\affiliation{\pnnl}
\affiliation{\usc}
\affiliation{\ornl}
\affiliation{\ITEP}
\affiliation{\JINR}
\affiliation{\duke}
\affiliation{\tunl}
\affiliation{\uw}
\affiliation{\unc}
\affiliation{\sdsmt}
\affiliation{\lanl}
\affiliation{\ut}
\affiliation{\ou}
\affiliation{\usd}
\affiliation{\princeton}
\affiliation{\ncsu}
\affiliation{\blhill}
\affiliation{\ttu}
\affiliation{\queens}

\author{N.~Abgrall}\affiliation{\lbnl}		
\author{I.J.~Arnquist}\affiliation{\pnnl} 
\author{F.T.~Avignone~III}\affiliation{\usc}\affiliation{\ornl}
\author{A.S.~Barabash}\affiliation{\ITEP}
\author{F.E.~Bertrand}\affiliation{\ornl}
\author{A.W.~Bradley}\affiliation{\lbnl}	
\author{V.~Brudanin}\affiliation{\JINR}
\author{M.~Busch}\affiliation{\duke}\affiliation{\tunl}	
\author{M.~Buuck}\affiliation{\uw}  
\author{T.S.~Caldwell}\affiliation{\unc}\affiliation{\tunl}	
\author{Y-D.~Chan}\affiliation{\lbnl}
\author{C.D.~Christofferson}\affiliation{\sdsmt} 
\author{P.-H.~Chu}\affiliation{\lanl} 
\author{C. Cuesta}\affiliation{\uw}	
\author{J.A.~Detwiler}\affiliation{\uw}	
\author{C. Dunagan}\affiliation{\sdsmt}	
\author{Yu.~Efremenko}\affiliation{\ut}
\author{H.~Ejiri}\affiliation{\ou}
\author{S.R.~Elliott}\affiliation{\lanl}
\author{T.~Gilliss}\affiliation{\unc}\affiliation{\tunl}  
\author{G.K.~Giovanetti}\affiliation{\princeton}  
\author{J. Goett}\affiliation{\lanl}	
\author{M.P.~Green}\affiliation{\ncsu}\affiliation{\tunl}\affiliation{\ornl}   
\author{J. Gruszko}\affiliation{\uw}		
\author{I.S.~Guinn}\affiliation{\uw}		
\author{V.E.~Guiseppe}\affiliation{\usc}
\author{C.R.S.~Haufe}\affiliation{\unc}\affiliation{\tunl}	
\author{R.~Henning}\affiliation{\unc}\affiliation{\tunl}
\author{E.W.~Hoppe}\affiliation{\pnnl}
\author{S.~Howard}\affiliation{\sdsmt}  
\author{M.A.~Howe}\affiliation{\unc}\affiliation{\tunl}
\author{B.R.~Jasinski}\affiliation{\usd}  
\author{K.J.~Keeter}\affiliation{\blhill}
\author{M.F.~Kidd}\affiliation{\ttu}	
\author{S.I.~Konovalov}\affiliation{\ITEP}
\author{R.T.~Kouzes}\affiliation{\pnnl}
\author{A.M.~Lopez}\affiliation{\ut}	
\author{J.~MacMullin}\affiliation{\unc}\affiliation{\tunl} 
\author{R.D.~Martin}\affiliation{\queens}	
\author{R. Massarczyk}\affiliation{\lanl}		
\author{S.J.~Meijer}\affiliation{\unc}\affiliation{\tunl}	
\author{S.~Mertens}\affiliation{\mpi}\affiliation{\tum}		
\author{C.~O'Shaughnessy}\affiliation{\unc}\affiliation{\tunl}	
\author{A.W.P.~Poon}\affiliation{\lbnl}
\author{D.C.~Radford}\affiliation{\ornl}
\author{J.~Rager}\affiliation{\unc}\affiliation{\tunl}	
\author{A.L.~Reine}\affiliation{\unc}\affiliation{\tunl}
\author{K.~Rielage}\affiliation{\lanl}
\author{R.G.H.~Robertson}\affiliation{\uw}
\author{B.~Shanks}\affiliation{\unc}\affiliation{\tunl}	
\author{M.~Shirchenko}\affiliation{\JINR}
\author{A.M.~Suriano}\affiliation{\sdsmt} 
\author{D.~Tedeschi}\affiliation{\usc}		
\author{J.E.~Trimble}\affiliation{\unc}\affiliation{\tunl}	
\author{R.L.~Varner}\affiliation{\ornl}  
\author{S. Vasilyev}\affiliation{\JINR}	
\author{K.~Vetter}\altaffiliation{Alternate address: Department of Nuclear Engineering, University of California, Berkeley, CA, USA}\affiliation{\lbnl}
\author{K.~Vorren}\altaffiliation{Corresponding author. Email: krisvorren@unc.edu}\affiliation{\unc}\affiliation{\tunl}
\author{B.R.~White}\affiliation{\lanl}	
\author{J.F.~Wilkerson}\affiliation{\unc}\affiliation{\tunl}\affiliation{\ornl}    
\author{C. Wiseman}\affiliation{\usc}		
\author{W.~Xu}\affiliation{\usd} 
\author{E.~Yakushev}\affiliation{\JINR}
\author{C.-H.~Yu}\affiliation{\ornl}
\author{V.~Yumatov}\affiliation{\ITEP}
\author{I.~Zhitnikov}\affiliation{\JINR}
\author{B.X.~Zhu}\affiliation{\lanl}
			
\collaboration{{The \textsc{Majorana}} Collaboration}
\noaffiliation
%
%
%
\vskip 0.25cm
\date{March 21, 2017}

\begin{abstract}
We present new limits on exotic keV-scale physics based on 478~kg~d of \mj~\dem\ commissioning data. Constraints at the 90\% confidence level are derived on bosonic dark matter (DM) and solar axion couplings, Pauli exclusion principle violating (PEPV) decay, and electron decay using monoenergetic peak signal-limits above our background. Our most stringent DM constraints are set for 11.8~keV mass particles, limiting $g_{Ae} <4.5\times 10^{-13}$ for pseudoscalars and $\frac{\alpha'}{\alpha} < 9.7\times 10^{-28}$ for vectors. We also report a 14.4~keV solar axion coupling limit of $g_{AN}^{\mathrm{eff}}\times g_{Ae}~<~3.8 \times 10^{-17}$, a $\frac{1}{2}\beta^2~<~8.5\times10^{-48}$ limit on the strength of PEPV electron transitions, and a lower limit on the electron lifetime of $\tau_e > 1.2 \times 10^{24}$~yr for $e^- \rightarrow$ invisible.
\end{abstract}

\pacs{95.35.+d}
\maketitle




The \mj\ \dem, described in detail in Ref.~\cite{MJD14}, is a neutrinoless double-beta decay (\nbb) experiment located 4850~ft underground at the Sanford Underground Research Facility in Lead, South Dakota~\cite{Heise2015}. \mj\ consists of two separate custom ultralow background modules, each containing seven arrays of $P$-type point contact (PPC) high-purity germanium (HPGe) detectors with a total mass of  44.1~kg, of which 29.7~kg is enriched to 88\% $^{76}$Ge.

The geometry of the PPC detectors results in low capacitance and reduced electronic noise, and permits good energy resolution with very low energy thresholds. In addition, PPC HPGe detectors have advantageous pulse-shape discrimination capabilities~\cite{luke89, Barbeau07, coop11}. Previous experiments have exploited these capabilities to perform high-sensitivity searches for light weakly interacting massive particles (WIMP) and bosonic dark matter (DM)~\cite{gio15, cogent14, Zhao2013} as well as \nbb\ decay searches~\cite{Klapdor01, IGEX02, Ackermann2013}.

In this Letter, we set limits on multiple keV-scale rare-event interactions from monoenergetic signal limits with 478~kg~d of \mj\ commissioning data. Bosonic pseudoscalar (i.e. axionlike) and vector DM, with mass scale of 1--100~keV, offer an explanation for the observed subgalactic structure in the Universe, assuming a large number density compensates for their light mass. With suitable electronic coupling strength, they may be detectable via a pseudoscalar or vector-electric effect that is analogous to photoelectric absorption~\cite{Pospelov08,Redondo2009,An2015331}. In addition, we report limits on the coupling of 14.4~keV solar axions competing in the $M$1 transition of~$^{57}$Fe nuclei, Pauli exclusion principle violating (PEPV) electronic transitions, and electron decay, $e^- \rightarrow$~invisible.



\mj\ relies on careful material selection and handling~\cite{abgr16} to reduce intrinsic and extrinsic radioactive background, making it well suited for dark matter and other rare-event searches. \mj\ modules are surrounded by a copper shield, a lead shield, an active muon veto~\cite{MJmuon}, and a polyethylene neutron shield. Within the shielding, radon is purged via liquid nitrogen boil-off. The inner 5~cm of the copper shield, the cryostats that house the detectors, and the crystal support structures are fabricated from radiopure ($<$0.1~$\mu$Bq/kg U) copper electroformed in an underground facility. 

\begin{figure}
\includegraphics[width=0.5\textwidth]{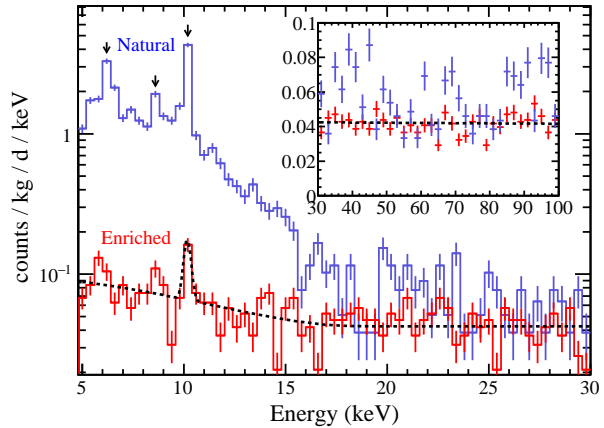}
\caption{\label{fig:cal} Energy spectra from 195~kg~d of natural (blue) and 478~kg~d of enriched (red) detector data. A fit of the background model (linear + tritium beta spectrum + $^{68}$Ge $K$ shell) to the enriched spectrum is also shown (dotted black). The background rate and slope, along with the tritium and $K$-shell rate were floated in the fit. The background fit $\chi^2/$NDF is 75.7/85. Cosmogenic isotopes in the natural detectors produce peaks at 10.36 ($^{68}$Ge), 8.9 ($^{65}$Zn), and 6.5~keV ($^{55}$Fe) on top of a tritium beta decay continuum. The FWHM of the 10.4~keV peak is $\sim$0.4~keV. The spectrum shown does not include a $T/E$ cut acceptance correction.}
\end{figure}

The data presented here were acquired during the June~30 to September~22, 2015 commissioning of \mj~Module~1 (M1). During this time, Module~2 was under construction and not operational. The shield was incomplete: the innermost 5~cm of electroformed copper shielding was not yet installed, the active muon-veto system was not finished, and the exterior neutron shielding did not fully enclose the inner layers. Shielding inside and outside the vacuum and cryogenic services still had to be added. 
The natural (unenriched) detectors had a high cosmogenic background compared to the enriched detectors because of different handling procedures, and were only used here for systematic studies; see Fig.~\ref{fig:cal}. Seven of the enriched detectors were inoperable due to failed electrical connections or high noise rates. The active mass of the remaining~13 enriched detectors was computed from detector dead layer measurements provided by ORTEC~\cite{ORTEC} and verified via collimated $^{133}$Ba source scans, totaling $10.06 \pm 0.13$ kg. The commissioning live time was $47.503\pm0.001$~d, resulting in an exposure of $478\pm6$~kg~d.

The data-acquisition (DAQ) system is controlled and monitored by the ORCA software package~\cite{How04}. Signals from the PPC detectors are amplified and shaped by a custom low-noise resistive-feedback preamplifier with a measured equivalent noise charge of $\sim$85~eV in Ge-detector-equivalent FWHM resolution~\cite{Barton2011}. The amplifier provides low-gain and high-gain outputs that are digitized separately by a custom 14-bit 100~MHz VME-based digitizer designed for the GRETINA experiment~\cite{pasc13}. Signals are digitized continuously and triggers are generated when the output of a firmware-based trapezoidal filter trigger exceeds the preset threshold for that channel.  An internal pulser ($\sim$0.1~Hz), implemented by injecting charge through capacitive coupling to the gate of the preamplifier's front-end JFET, is used to monitor detector live time and gain stability.

Transient and other irregular noise pulses from the DAQ hardware contaminate the energy spectrum between $2\,-\, 70$~keV.  Most of the nonphysical waveforms are due to accidental retriggering during baseline restoration after pulser events. These are removed by eliminating events with more than one detector hit or by using pulse-shape discrimination. The acceptance of these cuts is 99.98\% with negligible uncertainty.

Slow-pulse waveforms with rise times of $\sim$1~$\mu$s or longer constitute a significant background below 30~keV, as recognized by previous experiments \cite{Aguayo2013, cogent14, gio15, Zhao2013}. Slow pulses are energy-degraded events that originate in low-field regions of the detector near the surface dead layer, where diffusion is the dominant mode of charge transport. At energies $<$10~keV, discriminating slow pulses using pulse rise-time measurements becomes difficult since signal to noise ratio decreases with energy.

A more robust parameter, $T/E$, was developed to tag slow pulses. A trapezoidal filter with a 100~ns ramp time and a 10~ns flat-top time was applied to each waveform, and the maximum ($T$) value of the result was measured. The $T$ value was normalized by an energy parameter, ($E$), which was reconstructed offline by finding the maximum ~\footnote{Except for use in the $T/E$ parameter, the actual calculated energy is the filtered value at a fixed time of 6~$\mu$s after the start of the rising edge, identified using a second trapezoidal filter with a 1~$\mu$s ramp time and $\sim$1~keV threshold. The energy scale is also subject to a digitizer non-linearity correction.} of a trapezoidal filtered waveform with a filter rise time of 4~$\mu$s and flat-top time of 2.5~$\mu$s. This parameter exhibited good separation between fast and slow-pulse waveforms down to $\sim$3~keV, below the 5~keV analysis threshold.

The signal acceptance of the $T/E$ cut was measured by capacitively injecting simulated signal pulses of varying amplitude directly onto the detector's outer contact using a precision waveform generator. The energy dependent acceptance was determined by finding the fraction of these events that pass the cut at set pulse amplitudes. An error function was fit to the acceptance fractions to estimate the acceptance between pulser-peak events. Only three of the~13 analysis detectors were instrumented with the required electronics to perform this test and the smallest-valued (most conservative) acceptance function, ranging from 96\% at 5~keV to 100\% at 20~keV, was applied in the DM rate analysis, Eq.~\ref{eq:DMcounts}. The detector acceptance functions varied by at most 1\%. The energy dependent acceptance uncertainty was determined from the error function fit, 
\begin{equation}
\eta(E)\;=\frac{\mathrm{Erf}(E-\mu)}{\sqrt{2}\sigma}
\label{eq:ToEeff}
\end{equation}
The fit values were $\mu\;=\;-26\;\pm\;4$~keV and $\sigma\;=\;13.7\;\pm\;1.7$~keV with a strong anticorrelation, corr($\mu,\;\sigma)\;\sim\;-1$.



A $^{228}$Th line source inserted into a helical calibration track surrounding the cryostat was used for energy calibration. Multiple calibration periods were interspersed between background data collection to track and account for long-term drift in gain. Statistically significant peaks in the $^{228}$Th decay chain energy spectrum were used to calibrate the energy spectra of each detector independently. 
To extend our calibration to lower energies, we included the measured baseline noise as the zero point energy in the fit. For an overview of the calibration system, see~\cite{MJDcal}. 

We combined the calibration spectra from the 13 detectors, and summed a total of 102.8 hours of calibration data over all of the calibration periods. The resulting high statistics spectrum permitted peaks from Bi $x$ rays and from Th and Pb gamma rays. These were used to help quantify biases and uncertainties in the energy scale below 120~keV. A small systematic offset in the energy scale~($E_S$) of $\sim$0.2~keV from known peak energies was observed in this region. The offset is consistent with residual digitizer nonlinearity effects, which were estimated by comparing energy measurements from low-gain and high-gain channels. A linear correction ($\Delta E$),
\begin{equation}
 \label{eqn:linCorr}
\Delta E(E_S) = \alpha_E(E_S - 95.0\;\mathrm{keV}) + E_0\; ,
 \end{equation}
was applied to mitigate the offset. The parameters $\alpha_E = -0.0014 \pm 0.0008$ and~$E_0 = -0.256 \pm 0.016\; \mathrm{keV}$ were determined by fitting a line to the peak-centroid offset values of the low-statistics peaks between 70 and 120~keV. The correlation coefficient was corr$(\alpha_E, E_0) = -0.22$. 
The correction was then extrapolated to lower energies. As a check, the predicted offset at 10.36~keV, the $^{68}$Ge cosmogenic $K$-shell cascade peak, was computed and found to be $-0.12 \pm 0.07$~keV. In the natural detectors, this peak was measured at 10.22~keV, and is consistent with the correction model prediction in Eq.~\ref{eqn:linCorr} to within the parameter uncertainties. We are improving our nonlinearity correction and expect to remove this offset in future analyses.



A multipeak fitting routine was applied to the summed $^{228}$Th calibration spectrum to determine the energy dependent widths ($\sigma$) of peaks in the 1--260~keV energy range. 
The widths were fit to
\begin{equation}
\sigma_E(E) = \sqrt{\sigma_0^2 + \langle\varepsilon\rangle FE}\; ,
\label{eqn:resolution}
\end{equation}
%
%
with resulting fit values of $\sigma_0 = 0.16 \pm 0.04$~keV and $F = 0.11 \pm 0.02$. The fit parameters were fully correlated, corr$(\sigma_0, F)\sim1$. The constant~$\langle\varepsilon\rangle =$~2.96 eV, is the average energy required to produce an electron-hole pair in Ge. 

Limits on pseudoscalar dark matter axioelectric coupling were calculated using a method similar to~\cite{EDEL}. For comparison with other experiments, we set the Milky Way halo density to $\rho_{DM}=0.3$~GeV~cm$^{-3}$~\footnote{Recent results suggest that the local density is closer to 0.4~GeV~cm$^{-3}$. See Ref.~\cite{Hooper16}} and assumed that pseudoscalar DM constitutes the total density. The expected number of detected counts, $dN/dE$ at energy $E$, assuming a pseudoscalar mass of $m_A$ in keV, is given by~\cite{Aless12, EDEL}
\begin{align}
\frac{dN}{dE}&(E;m_A) = \Phi_{DM}(m_A)\sigma_{Ae}(m_A)\; \nonumber
\\
& \eta(E)\frac{1}{\sqrt{2\pi}\sigma_E(m_A)}\mathrm{exp}\left(-\frac{(E-m_A)^2}{2\sigma_E^2(m_A)}\right)MT \;,
\label{eq:DMcounts}
\end{align}
\begin{align}
&\Phi_{DM} = \rho_{DM}  \frac{v_A}{m_A} = 7.8 \times 10^{-4} \left(\frac{1}{m_A}\right)\cdot\beta\; [\mathrm{/barn/day}],
\label{eq:DMflux}
\\
&\sigma_{Ae}(m_A) = \sigma_{pe}(m_A)\frac{g_{Ae}^2}{\beta} \frac{3m_A^2}{16\pi\alpha m_e^2} \left( 1 - \frac{\beta^{\frac{2}{3}}}{3}\right).
\label{eq:DMcross}
\end{align}
where $\beta = v_A/c$ is the average DM velocity with respect to the earth, $\Phi_{\mathrm{DM}}$ is the average DM flux at Earth, $\sigma_{Ae}$ is the axioelectric cross section as a function of energy, $\sigma_E$ is the energy resolution at $E = m_A$ (given by Eq.~\ref{eqn:resolution}), $MT$ is the exposure of the detectors used in this analysis, and $\eta(E)$ is the $T/E$ cut acceptance function (Eq.~\ref{eq:ToEeff}). In Eq.~\ref{eq:DMcross}, $\sigma_{pe}$ is the photoelectric cross section in Ge~\cite{nistGe}. In this analysis, the peak energy of interest is the pseudoscalar mass ($m_A$).  We take $\beta = 0.001$ \cite{EDEL, XENON14}, roughly the mean of the dark matter velocity distribution with respect to Earth. 

We place an upper limit on the pseudoscalar dark matter coupling constant, $g_{Ae}$, at multiple $m_A$ values between 5 and 100~keV using an unbinned profile likelihood method~\cite{wilks1938,Rolke05,Eadie2006}. The likelihood function incorporates a DM signal probability density function that is modeled separately with Eq.~\ref{eq:DMcounts} for each individual $m_A$ value, a linear background, the tritium spectrum and a 10.36 keV cosmogenic $x$-ray peak. A multidimensional Gaussian penalty term floats the nuisance parameters ($\alpha_E$, $E_0$, $\sigma_E$, and $\eta$) in the likelihood function according to their covariance matrices. The penalty term affects the final limit by a few percent at most. 
The best fit to the background model is shown in Fig.~\ref{fig:cal}.

\begin{figure}[t]
\includegraphics[width=0.5\textwidth]{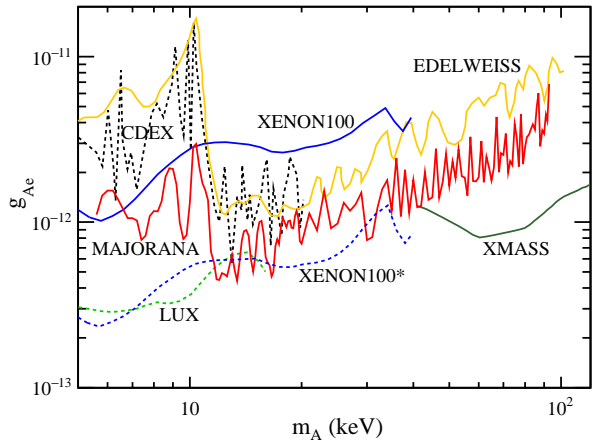}
\caption{\label{fig:psLimit} The 90\% UL on the pseudoscalar axionlike particle dark mater coupling from the \mj\ \dem\ (red) compared to EDELWEISS~\cite{EDEL} (orange), XMASS~\cite{XMASS14} (dark green), and XENON~\cite{XENON14} (blue). XENON has recently published an erratum~\cite{XenonEr} (dashed blue). Results by LUX (dashed, light green) have not yet been published~\cite{IDM16}, and new results from CDEX (dashed, black) are available in~\cite{CDEX16}.
}
\end{figure}

A comparison of our $g_{Ae}$-limits, as a function of pseudoscalar mass, to previous results is shown in Fig.~\ref{fig:psLimit}. 
Our limits are an improvement over other germanium experiments, EDELWEISS~\cite{EDEL} and CDEX~\cite{CDEX16}, especially for $m_A<18.6$~keV due to the low cosmogenic activity in \mj\ enriched detectors. The XMASS~\cite{XMASS14} experiment has the best limits for~$m_A>40$~keV. Two XENON limits are shown: the original published in~\cite{XENON14} (solid), and a correction from an erratum~\cite{XenonEr} (dashed). Preliminary LUX results~\cite{IDM16} are comparable to the revised XENON results. Currently the xenon experiments XMASS, XENON, and LUX report the best limits due to the $>$10$\times$ larger exposure of their fiducial mass.

\begin{figure}[h]
\includegraphics[width=0.5\textwidth]{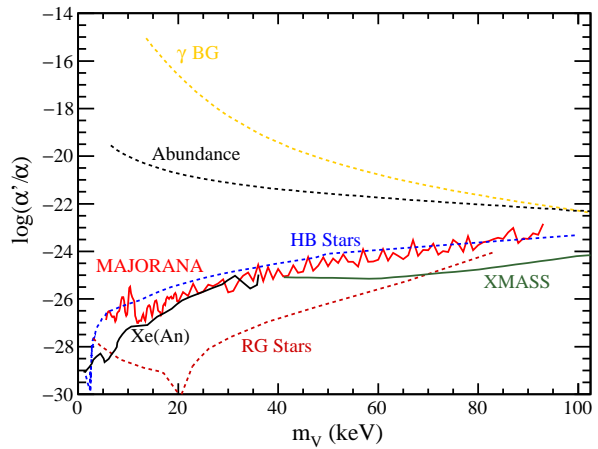}
\caption{\label{fig:vecLimit}The 90\% UL on the vector particle dark matter coupling from the \mj\ \dem\ (red) compared to the astrophysical limits (dashed) from the gamma background (orange), the observed dark matter abundance (black), horizontal branch stars (blue), and RG stars (maroon)~\cite{Pospelov08, Essig:2013lka}. Experimental results (solid) from XMASS~\cite{XMASS14} (green) along with a $2\sigma$ limit computed from XENON100~\cite{XENON14} data by H. An $et al.$~\cite{An2015331} are also shown.} 
\end{figure}

Using the same data and analysis technique with a Gaussian modeled signal, we also set limits on the electronic coupling of vector bosonic DM ~\cite{Pospelov08}. The interaction rate for vector DM is
\begin{equation}
\Phi_{DM}(m_V)\sigma_{Ve}(m_V)= \frac{4 \times 10^{23}}{m_V}\left(\frac{\alpha'}{\alpha}\right)\frac{\sigma_{pe}(m_V)}{A}\; [\mathrm{/kg /d}], 
\label{eq:vecDM}
\end{equation}
 where $A$ is the atomic mass of Ge, $m_V$ is the vector boson mass in keV, and $\alpha'$ is the coupling of vector DM to electrons, analogous to the electromagnetic fine structure constant, $\alpha$. The expected number of detector counts at energy $E$ is found by replacing the axioelectric interaction rate in Eq.~\ref{eq:DMcounts} with the vector-electric rate, with $m_V$ substituted for $m_A$. Limits on the vector coupling from the unbinned likelihood analysis described above are shown in Fig.~\ref{fig:vecLimit}. In the case of vector DM, the experimental constraints are more stringent than astrophysical limits, except for red giant (RG) stars.

\begin{figure}[h]
\includegraphics[width=0.5\textwidth]{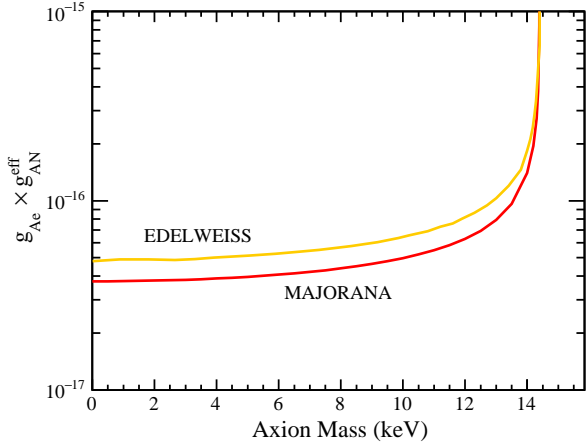}
\caption{\label{fig:hadronic}The 90\% UL coupling of 14.4-keV solar axions from the \mj\ \dem\  (red) data compared with the limit set by EDELWEISS (orange). The product of the axionuclear coupling in the sun and the axioelectric coupling in the detector is shown. Comparative astrophysical limits assuming $g_{AN}^{\mathrm{eff}}$ follows the DFSZ model is shown in Ref.~\cite{EDEL}.}
\end{figure}

In addition to generic pseudoscalar and vector DM, we analyzed our sensitivity to solar axions. $^{57}$Fe has a large solar abundance and its first excited state at  14.4~keV is thermally excited within the Sun's interior. Axion emission is possible from the decay of this state~\cite{Moriyama1995}. Electric coupling of these axions to atomic electrons in the detector would manifest as a peak at 14.4~keV. No such peak was observed in \mj, and a limit on the product of the effective axionuclear coupling, $g_{AN}^{\mathrm{eff}}$, of solar axions (see \cite{Andria09}) and the axioelectric coupling, $g_{Ae}$, was determined. Replacing the flux in Eq.~\ref{eq:DMflux} with~\cite{EDEL}
\begin{equation}
\Phi_{14.4} = \beta^3 \times 4.56 \times 10^{23} (g_{AN}^{\mathrm{eff}})^2\; [\mathrm{/cm^{2} /s}],
\label{eq:flux14.4}
\end{equation}
and substituting $m_A$ in Eq.~\ref{eq:DMcounts} with 14.4~keV, we use the unbinned likelihood analysis to determine a limit on the coupling constant. Since this is a monoenergetic transition, the reduced axion velocity, $\beta$, depends on the mass of the axion, which can range from 0 to 14.4~keV. In the low mass limit where $\beta\rightarrow 1$, we find a 90\% UL of $g_{AN}^{\mathrm{eff}}\times g_{Ae}< 3.8\times 10^{-17}$. A comparison of the \mj\ and EDELWEISS coupling limits is shown in Fig.~\ref{fig:hadronic}.

Two other non-DM related rare-event searches were carried out using the low energy data and analysis, a Pauli exclusion violating decay, and an electron decay search. While the Pauli exclusion principle is a fundamental law of nature, its physical origin is still not fully understood~\cite{Ignatiev1987,Greenberg1987,Greenberg1991,Greenberg2000,Elliott2012, Abgrall2016}. \mj\ searched for the PEPV transition of an $L$-shell Ge electron to the $K$ shell that would manifest as a 10.6~keV~\cite{Elliott2012} shoulder on the 10.36~keV $^{68}$Ge peak. Using the unbinned likelihood method with a generic signal plus background model, we set a 90\% C.L. on the excess signal rate of 0.03 /kg/d. This equates to a lifetime $\tau > 2.0 \times 10^{31}$ s. Comparing to the $1.7\times10^{-16}$ s lifetime of a standard $K_{\alpha}$ transition in Ge, one derives an upper limit on the PEPV parameter $\frac{1}{2}\hat\beta^2 < 8.5\times10^{-48}$, a $\sim$35\% improvement over the previous limit~\cite{Bernabei2009}.

Our data can also be used to set a limit on the decay of the electron. Charge conservation arises from an exact gauge symmetry of quantum electrodynamics with the associated gauge boson being exactly massless. Even so, the possibility of its violation has been theoretically explored~\cite{Okun1978,Voloshin1978a,Ignatiev1978,Mohapatra1987,Okun1989,Mohapatra1992,Ignatiev1996}. For example, the charge-conservation violating process $e^- \rightarrow \nu \bar{\nu} \nu$ produces an atomic-shell hole. If an electron disappears from the $K$~shell of a Ge atom, resulting atomic emissions deposits 11.1~keV of energy within the detector.  We search for events of this characteristic energy as possible indications of electron decay using a similar analysis as for the PEPV and solar axion search.  We determined a lifetime limit of $>$1.2$\times10^{24}$~yr. The best limit on the lifetime for this process is $>$2.4$\times10^{24}$~yr (90\% CL)~\cite{Belli1999}. 

We found no indication of new physics that would manifest as a peak in the energy spectrum of the Module~1 commissioning data presented in this Letter. Upgrades to \mj, detector repairs, and the addition of Module~2 will significantly improve the sensitivity to new physics. Lower background rates in subsequent data sets have already been observed with the installation of the inner electroformed copper and additional polyethylene neutron shielding. Analysis thresholds below 5~keV will allow us to constrain additional processes including light-WIMP scattering. 

This material is based upon work supported by the U.S. Department of Energy, Office of Science, Office of Nuclear Physics under Award  Numbers DE-AC02-05CH11231, DE-AC52-06NA25396, DE-FG02-97ER41041, DE-FG02-97ER41033, DE-FG02-97ER41042, DE-SC0012612, DE-FG02-10ER41715, DE-SC0010254, and DE-FG02-97ER41020. We acknowledge support from the Particle Astrophysics Program and Nuclear Physics Program of the National Science Foundation through grant numbers PHY-0919270, PHY-1003940, 0855314, PHY-1202950, MRI 0923142 and 1003399. We acknowledge support from the Russian Foundation for Basic Research, grant No. 15-02-02919. We  acknowledge the support of the U.S. Department of Energy through the Los Alamos National Lab/Laboratory Directed Research and Development Program. This research used resources of the Oak Ridge Leadership Computing Facility, which is a DOE Office of Science User Facility supported under Contract DE-AC05-00OR22725. This research used resources of the National Energy Research Scientific Computing Center, a DOE Office of Science User Facility supported under Contract No. DE-AC02-05CH11231. We thank our hosts and colleagues at the Sanford Underground Research Facility for their support.

\bibliography{bosonDM}
\end{document}